\documentclass[a4paper,11pt]{article}
\usepackage{jheppub} 
\usepackage[T1]{fontenc} 
\RequirePackage[displaymath]{lineno} % Display line numbers
\usepackage{graphicx}% Include figure files
\usepackage{dcolumn}% Align table columns on decimal point
\usepackage{bm}% bold math
\usepackage{ltablex,booktabs}
\usepackage{overpic}
\usepackage{subfigure}
\usepackage{float}
\usepackage{color}
\usepackage{amsmath}
\usepackage{mathcomp}
\usepackage{mathrsfs}
\usepackage{multirow}
\usepackage{amssymb}
\usepackage{gensymb}
\usepackage{amsmath}
\usepackage{tabularx}
\usepackage{epstopdf}
\usepackage{ulem}

\title{Search for lepton number violating decays of $D_s^+\to h^-h^0e^+e^+$}

\collaborationImg{\includegraphics[width=0.15\textwidth, angle=90]{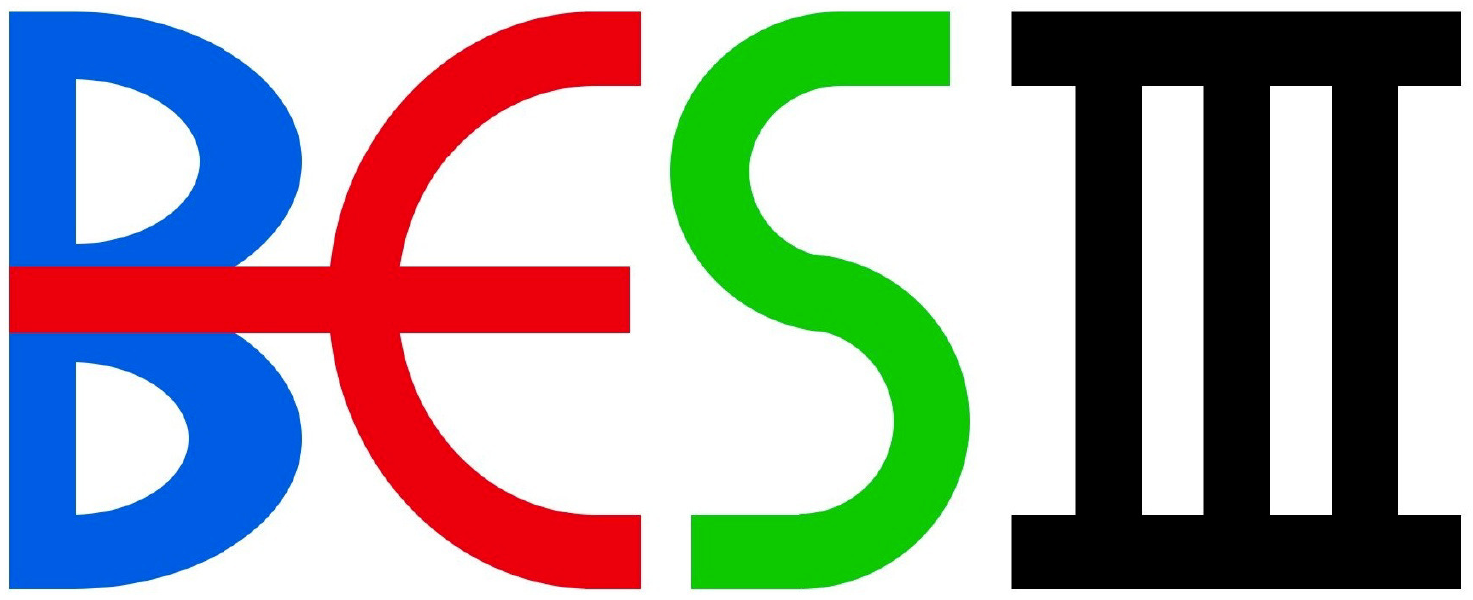}}

\collaboration{The BESIII Collaboration}

\emailAdd{besiii-publications@ihep.ac.cn}

\abstract{Based on 7.33 fb$^{-1}$ of $e^+e^-$ collision data collected by the BESIII detector operating at the BEPCII collider at center-of-mass energies from 4.128 to 4.226 GeV, a search for the Majorana neutrino $\nu_m$ is conducted in the lepton-number-violating decays of $D_s^+\to h^-h^0e^+e^+$. Here, $h^-$ represents a $K^-$ or $\pi^-$, and $h^0$ represents a $\pi^0$, $K_S^0$ or $\phi$. No significant signal is observed, and the upper limits of their branching fractions at the 90\% confidence level are determined to be $\mathcal{B}(D_s^+\to \phi\pi^-e^+e^+) < 6.9 \times 10^{-5}$, $\mathcal{B}(D_s^+\to \phi K^-e^+e^+) < 9.9 \times 10^{-5}$, $\mathcal{B}(D_s^+\to K_S^0\pi^-e^+e^+) < 1.3 \times 10^{-5}$, $\mathcal{B}(D_s^+\to K_S^0K^-e^+e^+) < 2.9 \times 10^{-5}$, $\mathcal{B}(D_s^+\to \pi^-\pi^0e^+e^+) < 2.9 \times 10^{-5}$ and $\mathcal{B}(D_s^+\to K^-\pi^0e^+e^+) < 3.4 \times 10^{-5}$. The Majorana neutrino is searched for with different mass assumptions within the range [0.20, 0.80] GeV$/c^2$ in the decay of $D_s^+\to\phi e^+\nu_m$ with $\nu_m\to\pi^-e^+$, and the upper limits of the branching fractions at the 90\% confidence level are at the level of $10^{-5}-10^{-2}$, depending on the mass of the Majorana neutrino.}

    
\arxivnumber{2410.02421}

\begin{document}
%\linenumbers
\maketitle
\flushbottom

\newpage
\section{Introduction}
%\label{int}

Neutrinos are expected to be massless in the Standard Model (SM). However, the discovery of neutrino oscillations~\cite{nu_os1,nu_os2,nu_os3} and the observation of the $\theta_{13}$ mixing angle~\cite{theta13} have convincingly shown that neutrinos have mass. Nevertheless, whether neutrinos are Dirac or Majorana particles is still an open question. A popular model that naturally generates light neutrino masses is the so-called ``see-saw'' mechanism~\cite{see-saw1,see-saw2,see-saw3,see-saw4}, in which the small value of the observed neutrino masses arises from the existence of a heavy Majorana neutrino state with a mass from a few hundred MeV to a few GeV. Assuming the see-saw mechanism underlies the neutrino mass generation, the light neutrino in the large-small-mass Majorana neutrino pair allows processes that violate lepton number by two units ($\Delta L=2$). The most promising way to identify the nature of neutrinos is searching for neutrinoless double beta $(0\nu\beta\beta)$ nuclear decay~\cite{beta1,beta2,beta3,beta4,beta5}. So far, only limits on its rates have been obtained, allowing bounds to be set on the effective Majorana mass at the level of $10^{-1}$ eV~\cite{beta_sum}.

Among all possible $\Delta L=2$ processes, interesting sources of lepton-number-violating (LNV) reactions are characterised by the exchange of a single Majorana neutrino whose mass is on the scale of heavy flavor mass~\cite{dL2}. This heavy Majorana neutrino can be on-shell, enhancing its contribution~\cite{dL2}. The BESIII experement has searched for four-body $\Delta L=2$ decays of $D\to K \pi e^+ e^+$, with measured upper limits (ULs) of branching fractions (BFs) at $10^{-6}$ level~\cite{DKpiee}. The LHCb experiment has widely studied three-body $\Delta L=2$ decays of $D^+$ and $D_s^+$ mesons, with measured ULs of BFs around $10^{-8}-10^{-6}$~\cite{threebody}. However, according to the latest experimental results of ULs at the 90\% confidence level (C.L.) for $D_s^+$ decays compiled by HFLAV~\cite{HFLAV}, there is still no result for four-body $\Delta L=2$ decays. Some models predict the BFs of four-body $\Delta L=2$ charm meson decays to be up to $\mathcal{O}(10^{-6})$, potentially within reach of current experimental data~\cite{dL2, fourbody1, Dhhee, fourbody2}. 

\begin{figure}
    \centering
    \subfigure[$D_s^+\to \phi\pi^-e^+e^+$ (CF)]{
    \includegraphics[width=4.5cm]{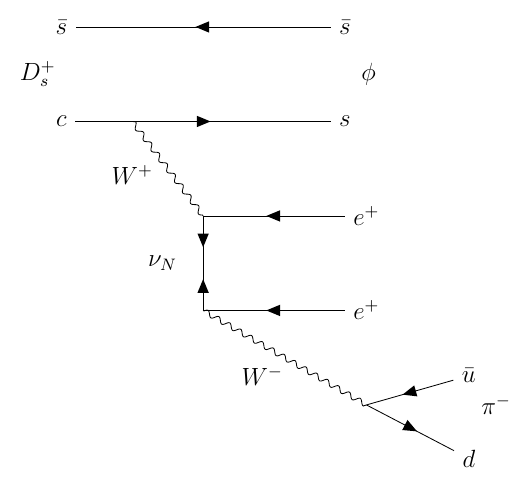}}
    \subfigure[$D_s^+\to \phi K^-e^+e^+$ (SCS)]{
    \includegraphics[width=4.5cm]{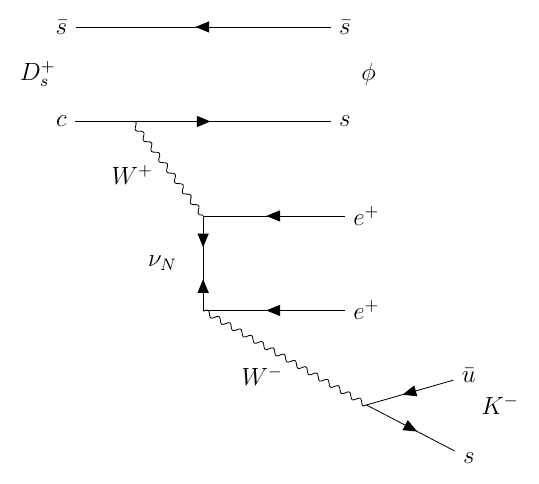}}
    \subfigure[$D_s^+\to K_S^0\pi^-e^+e^+$ (SCS)]{
    \includegraphics[width=4.5cm]{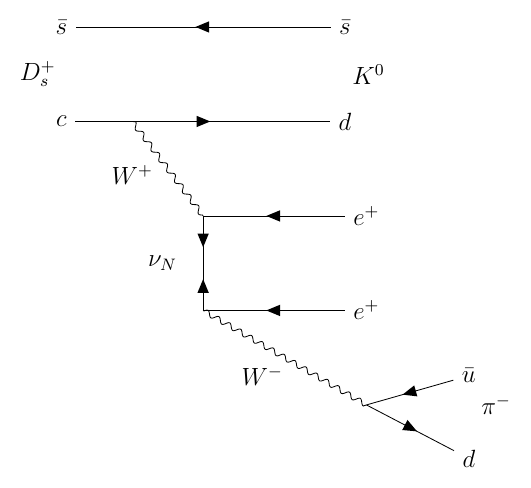}}
    \subfigure[$D_s^+\to K_S^0K^-e^+e^+$ (DCS)]{
    \includegraphics[width=4.5cm]{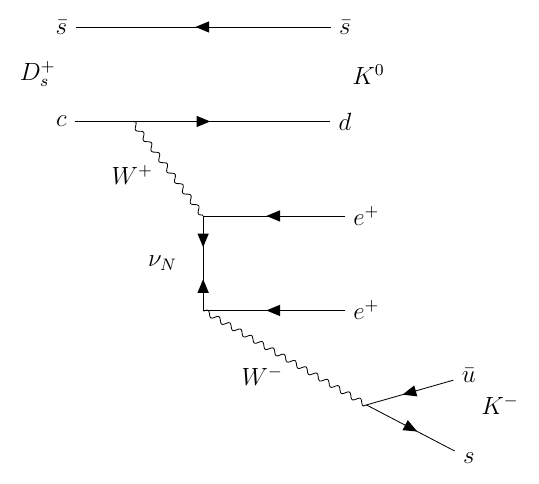}}
    \subfigure[\mbox{$D_s^+\to \pi^-\pi^0e^+e^+$ ($W$-exc.)}]{
    \includegraphics[width=4.5cm]{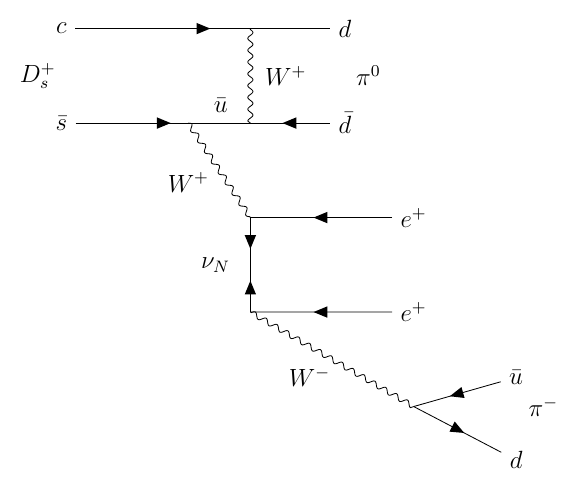}}
    \subfigure[\mbox{$D_s^+\to K^-\pi^0e^+e^+$ ($W$-exc.)}]{
    \includegraphics[width=4.5cm]{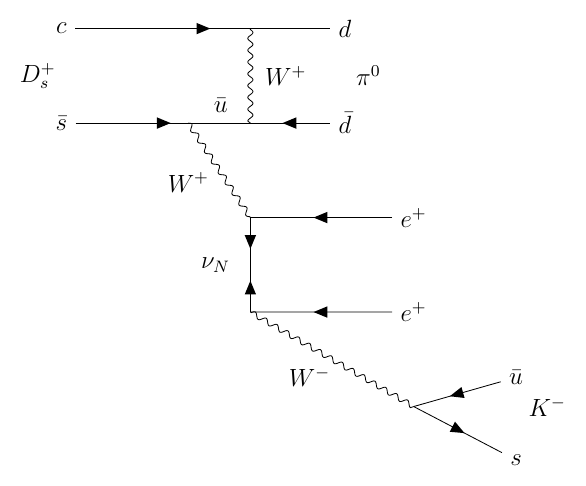}}
    \caption{Feynman diagrams for four-body LNV decays of $D_s^+$, where SCS represents singly Cabibbo suppressed, DCS represents doubly Cabibbo suppressed, and $W$-exc. represents $W$-exchange.}
    \label{feynman}
\end{figure}

In this work, a search for four-body LNV decays $D_s^+\to h^-h^0e^+e^+$ is performed based on a 7.33 fb$^{-1}$ data sample taken at center-of-mass~(c.m.) energies~($\sqrt{s}$) from 4.128 to 4.226 GeV collected by the BESIII experiment. Here $h^-$ represents a $K^-$ or $\pi^-$, and $h^0$ represents a $\pi^0$, $K_S^0$ or $\phi$. Figure~\ref{feynman} shows the Feynman diagrams for these processes. For the Cabibbo favored (CF) decay $D_s^+\to \phi\pi^-e^+e^+$, the ULs of the BF based on different mass assumptions of the Majorana neutrino are measured. In the following text, the mass of the Majorana neutrinos will be referred to as $m_{\nu_m}$. The other five decays are either Cabibbo-suppressed or involve W-exchange, leading to significantly reduced contributions compared to the CF decay $D_s^+\to\phi\pi^-e^+e^+$. As a result, we do not perform measurements of the ULs of the BF based on different $m_{\nu_m}$ assumptions for these decays. Throughout this paper, the charged conjugated modes are implied.

\section{Detector and data sets}
%\label{detector_dataset}

The BESIII detector~\cite{BESIII:2009fln} records symmetric $e^+e^-$ collisions provided by the BEPCII storage ring~\cite{Yu:2016cof}, in c.m. energies range from 1.84 to 4.95 GeV, with a peak luminosity of \mbox{$1.1\times 10^{33}$ cm$^{-2}$s$^{-1}$} achieved at the c.m. energy of 3.773 GeV. BESIII has collected large data samples in this energy region~\cite{Ablikim:2019hff}. The cylindrical core of the BESIII detector covers 93\% of the full solid angle and consists of a helium-based multilayer drift chamber~(MDC), a plastic scintillator time-of-flight system~(TOF), and a CsI(Tl) electromagnetic calorimeter~(EMC), which are all enclosed in a superconducting solenoidal magnet providing a 1.0~T magnetic field~\cite{Huang:2022wuo}. The solenoid is supported by an octagonal flux-return yoke with resistive plate counter muon identification modules interleaved with steel. The charged-particle momentum resolution at 1~GeV/$c$ is $0.5\%$, and the ${\rm d}E/{\rm d}x$ resolution is $6\%$ for electrons from Bhabha scattering. The EMC measures photon energies with a resolution of $2.5\%$ ($5\%$) at 1~GeV in the barrel (end cap) region. The time resolution in the TOF barrel region is 68~ps, while that in the end cap region is 110~ps. The end cap TOF system was upgraded in 2015 using multi-gap resistive plate chamber technology, providing a time resolution of 60~ps, which benefits 83\% of the data used in this analysis~\cite{etof1,etof2,etof3}. 

The data samples are organized into four groups, ${\sqrt{s} = 4.128}$ and 4.157 GeV, 4.178 GeV, four energies from 4.189 to 4.219 GeV (labelled as ``$4.189-4.219$''), and ${4.226~\rm{GeV}}$, acquired during the same year under consistent running conditions. The integrated luminosities at each energy are given in Table~\ref{energe}. Since the production cross section of $e^+e^-\to D^{*\pm}_s D^\mp_s$ in $e^+e^-$ annihilation is about twenty times larger than the $e^+e^-\to D^+_sD^-_s$ process~\cite{CLEO:2008ojp}, 
the signal events discussed in this paper are selected from the $e^+e^-\to D^{*\pm}_s D^\mp_s$ process.

\begin{table}[htbp]
 \centering
 \caption{The integrated luminosities ($\mathcal{L}_{\rm int}$) for various c.m. energies. The first and second uncertainties are statistical and systematic, respectively. The integrated luminosities for data samples of ${\sqrt{s} = 4.128}$ and 4.157 GeV are estimated using online monitoring information.}
 \begin{tabular}{l c c}
 \hline
      Category  & $\sqrt{s}$ (GeV) &$\mathcal{L}_{\rm int}$ (pb$^{-1}$)~\cite{luminosities1,luminosities2} \\
 \hline
	 \multirow{2}{*}{4.128\&4.157}   & 4.128 &401.5 \\
	                               & 4.157 &408.7 \\\hline
	 4.178                           & 4.178 &$3189.0\pm0.2\pm31.9$ \\\hline
	 \multirow{4}{*}{$4.189-4.219$}  & 4.189 &$570.0\pm0.1\pm2.2$ \\
	                                 & 4.199 &$526.0\pm0.1\pm2.1$ \\
	                               & 4.209 &$572.1\pm0.1\pm1.8$ \\
	                               & 4.219 &$569.2\pm0.1\pm1.8$ \\\hline
	 4.226                           & 4.226 &$1100.9\pm0.1\pm7.0$ \\
  \hline
 \end{tabular}
 \label{energe}
\end{table}

Simulated samples produced with the {\sc geant4}-based~\cite{GEANT4:2002zbu} Monte Carlo (MC) program, which includes the geometric description of the BESIII detector and the detector response, are used to determine the detection efficiency and estimate backgrounds. The samples include open charm processes, the initial-state radiation production of vector charmonium(-like) states and the continuum processes incorporated in {\sc kkmc}~\cite{Jadach:2000ir, Jadach:1999vf}. The known decay modes of charmed hadrons are modeled using {\sc evtgen}~\cite{Lange:2001uf, EVTGEN2} with BFs quoted from the Particle Data Group (PDG)~\cite{PDG}, otherwise estimated with {\sc lundcharm}~\cite{Chen:2000tv, LUNDCHARM2}. Final state radiation from charged final state particles is incorporated using {\sc photos}~\cite{PHOTOS}. The signal detection efficiencies and signal shapes are obtained with the signal MC samples. The signal processes are generated using a phase space (PHSP) model employed in {\sc evtgen}~\cite{Lange:2001uf, EVTGEN2}. The subsequent decays of $K_{S}^{0}\to \pi^+\pi^-$ and $\pi^0\to\gamma\gamma$ are also modeled using PHSP, while the $\phi\to K^{+}K^{-}$ decay is described by the VSS model, which parameterizes the decay of a vector meson to a pair of scalar particles. For the $D_s^+\to\phi e^+\nu_m(\to\pi^-e^+)$ process, signal MC samples are produced with varying assumptions for the Majorana neutrino mass, $m_{\nu_m}$. The modulus square of the amplitude for these processes is calculated based on previous theoretical work~\cite{Dhhee}.

\section{Methodology}
%\label{method}

Considering the expected small BFs and low background levels, a single tag (ST) method similar to that used in Ref.~\cite{DsFCNC} is applied. This method requires only one $D_s^+$ meson to be fully reconstructed in the signal mode for each event. The BF of $D_s^+\to h^-h^0e^+e^+$ decays can be calculated by
\begin{equation}
    \mathcal{B}(D_s^+\to h^-h^0e^+e^+) = \frac{N_{\rm sig}}{2\cdot N_{D^{*\pm}_s D^\mp_s}\cdot\epsilon\cdot \mathcal{B}_{\rm inter}},
\end{equation}
where $N_{\rm sig}$ is the signal yield, ${N_{D^{*\pm}_s D^\mp_s}=(64.72\pm 0.28)\times 10^5}$ is the total number of $D^{*\pm}_s D^\mp_s$ pairs in the data samples~\cite{Ds_number} and $\epsilon$ is the signal efficiency, obtained from the corresponding MC simulation, weighted over eight energy points given by ${\epsilon=\sum_{i=1}^8 \epsilon^iN^i_{D^{*\pm}_s D^\mp_s}/N_{D^{*\pm}_s D^\mp_s}}$, where $\epsilon^i$ and $N^i_{D^{*\pm}_s D^\mp_s}$ are the detection efficiency and the number of $D^{*\pm}_s D^\mp_s$ pairs at the $i$-th energy point, respectively. Finally, $\mathcal{B}_{\rm inter}$ is the BF of intermediate state decay.

\section{Event selection}
%\label{event_select}

Charged tracks are reconstructed from hits in the MDC and are required to be within a polar angle ($\theta$) range such that ${|\!\cos(\theta)|<0.93}$, where $\theta$ is defined with respect to the $z$ axis, which is the symmetry axis of the cylindrical MDC. For charged tracks not originating from $K_S^0$ decays, the distance of closest approach to the interaction point~(IP) must be less than 10~cm along the $z$ axis, $|V_z|$, and less than 1~cm in the transverse plane, $|V_{xy}|$. The momentum of charged tracks measured in the MDC must be greater than 100 MeV/$c$ to suppress the background from slow pions from $D^{*-}$ decays. Particle identification~(PID) for charged tracks combines the measurements of the specific ionization energy loss (${\rm d}E/{\rm d}x$) in the MDC and the flight time information in the TOF to form likelihoods $\mathcal{L}(h)~(h=K,\pi)$ for each hadron $h$ hypothesis, and a $K(\pi)$ is identified by requiring ${\mathcal{L}(K)>\mathcal{L}(\pi)}$ (${\mathcal{L}(\pi)>\mathcal{L}(K)}$). 

The $e^+$ candidate is required to have deposited energy greater than 25 MeV in the EMC. Its deposited energy in the EMC is required to be greater than 0.8 $c$ (0.7 $c$) times the measured momentum in the MDC for $e^+$ with momentum larger (smaller) than 400 MeV/$c$. This requirement is not applied to $D_s^+\to\phi K^-e^+e^+$ decay, but  is mandatory for  at least one $e^+$ in the $D_s^+\to\phi\pi^-e^+e^+$ decay and for both $e^+$ in other signal modes. This differentiation is necessary because signal modes with a narrower phase space typically produce electrons with lower momenta, which, due to the detector design, leads to lower tracking efficiencies and a reduced ability of the EMC to develop showers. Consequently, this leads to a lower signal efficiency. To address this issue, different requirements are applied for $e^+$ selections for different signal channels. PID for positrons uses the ${\rm d}E/{\rm d}x$, TOF and EMC information to construct likelihoods for positron, pion and kaon hypotheses ($\mathcal{L}(e), \mathcal{L}(\pi)$ and $\mathcal{L}(K)$). For the $D_s^+\to\phi K^-e^+e^+$ decay, at least one $e^+$ must meet the criterion ${\mathcal{L}(e)/(\mathcal{L}(e)+\mathcal{L}(\pi)+\mathcal{L}(K))>0.8}$, while for other signal modes, and both $e^+$ candidates are required to satisfy this criterion. The efficiency for the $D_s^+\to\phi K^-e^+e^+$ decay is found to be significantly lower than that for other signal channels, thus the PID requirement for $e^+$ is further loosened for this channel. Furthermore, photons within a cone of $5^{\circ}$ around the $e^+$ direction are recovered to the $e^+$ momentum to improve the momentum resolution.

The $K_{S}^0$ candidate is reconstructed from two oppositely charged tracks that satisfy \mbox{$|V_z|<20$~cm}. The two charged tracks are assigned as $\pi^{+}\pi^{-}$ without additional PID criteria. They are constrained to originate from a common vertex and are required to have an invariant mass ($M_{\pi^{+}\pi^{-}}$) within ${[0.487, 0.511]~\rm{GeV}/c^{2}}$. 

Photon candidates are identified using showers in the EMC. The deposited energy of each shower must be greater than 25~MeV in the barrel region~(${|\!\cos (\theta)|< 0.80}$) and more than 50~MeV in the end cap region~(${0.86 <|\!\cos (\theta)|< 0.92}$). To exclude showers originating from charged tracks, the minimum opening angle between the momentum of the candidate shower and the extrapolated direction on the EMC of any charged track must be greater than $10^{\circ}$. The difference between the EMC time and the event start time is required to be within \mbox{[0, 700]~ns} to suppress electronic noise and showers unrelated to the event. 

The $\pi^0$ candidates are reconstructed from photon pairs with invariant masses in the ranges ${[0.115, 0.150]~\rm{GeV}/c^{2}}$. A kinematic fit, which constrains the $\gamma\gamma$ invariant mass to the known $\pi^0$ mass~\cite{PDG}, is performed to improve the $\pi^0$ mass resolution. 

The $\phi$ candidates are reconstructed from $K^+K^-$ pairs with invariant masses in the ranges ${[1.00, 1.05]~\rm{GeV}/c^{2}}$. For $D_s^+\to\phi K^-e^+e^+$ decay, the $K^+K^-$ pair with an invariant mass closest to the known $\phi$ mass~\cite{PDG} is selected from the two possible combinations.

\section{Background analysis}
Background contributions are studied using inclusive MC samples. A variable, cos($\theta_{e^+e^-}$), is defined to be the cosine of the minimum angle between the first signal positron tagged and the oppositely charged track identified as an electron on the signal side. To suppress backgrounds from gamma conversion, $q\Bar{q}$ and Bhabha processes, cos($\theta_{e^+e^-}$) is required to be smaller than 0.95. For signal modes involving a charged pion, the cosine of the angle between $e^+$ and $\pi^-$~(cos($\theta_{e^+\pi^-}$)) is required to be smaller than 0.98 for both $e^+$ to suppress the backgrounds from $e^\pm/\pi^\pm$ misidentification. For signal modes involving a charged pion but without $K_{S}^0$, a veto is applied to $\pi^\pm$ originating from $K_{S}^0\to \pi^{+}\pi^{-}$ decays rather than from direct $D_s^{\pm}$ decays by requiring $L_\pi/\sigma_{L_\pi}<3$, where $L_\pi$ is the distance from the vertex of any $\pi^+\pi^-$ combination to the IP and $\sigma_{L_\pi}$ is the corresponding error. For signal modes with $K_{S}^0$, a criterion of ${L/\sigma_L>2}$ is applied to exclude fake $K_{S}^0$ candidates, where $L$ and $\sigma_L$ are the decay length of the $K_{S}^0$ candidate and its error, respectively. For signal modes with $\pi^0$, the energy of $\pi^0$ candidates~($E(\pi^0)$) is required to be greater than 0.17~GeV to remove soft $\pi^0$ from $D^{*}$ decays. 

A simultaneous requirement on $E/p$ and $\chi_{{\rm d}E/{\rm d}x}^2$ of $e^\pm$ is performed for the four signal channels without $\phi$, where $E$, $p$ and $\chi_{{\rm d}E/{\rm d}x}^2$ are the deposited energy in the EMC, the measured momentum in the MDC and $\chi^2$ of the specific energy loss, respectively. These requirements are determined independently for each mode based on the Punzi figure-of-merit $\epsilon/(\frac{3}{2}+\sqrt{B})$~\cite{Punzi}, where $\epsilon$ is the signal efficiency and $B$ is the number of background events. The optimized requirements on $E/p$ and $\chi_{{\rm d}E/{\rm d}x}^2$ for the four signal channels are listed in Table~\ref{eopchi2}.

\begin{table}[h]
\centering
\caption{Requirements on $E/p$ and $\chi_{{\rm d}E/{\rm d}x}^2$.}
\label{eopchi2}
\tabcolsep=0.2cm
\begin{tabular}{lcc}
\hline
Decay channel & $E/p$ & $\chi_{{\rm d}E/{\rm d}x}^2$\\\hline
$D_s^+\to K_S^0\pi^-e^+e^+$ & $>0.74~c$ & $<5.7$\\
$D_s^+\to K_S^0K^- e^+e^+$ & $>0.70~c$ & $<5.6$\\
$D_s^+\to \pi^-\pi^0e^+e^+$ & $>0.81~c$ & $<4.3$\\
$D_s^+\to K^-\pi^0e^+e^+$ & $>0.75~c$ & $<5.0$\\
\hline
\end{tabular}
\end{table}

To further suppress backgrounds and identify the $D_s^+$ candidates from $e^+e^-\to D^{*\pm}_s D^\mp_s$, the recoiling mass of $D^+_s$ ($M_{\rm rec}$) and the mass difference ($\Delta M$) are defined as:
\begin{equation}
\begin{aligned}
    &M_{\rm rec} = \sqrt{\left(E_{\rm cm}/c^2 - \sqrt{|\vec{p}_{D^+_s}|^2/c^2 + m^2_{D^+_s}}\right)^2 - |p_{D^+_s}|^2/c^2}, \\
    &\Delta M = M(D^+_s\gamma) - M(D^+_s),
\end{aligned}
\end{equation}
where $E_{\rm cm}$, $p_{D^+_s}$ and $m_{D_s^+}$ are the energy of $e^+e^-$ system, the three-momentum of $D^+_s$ candidate in the $e^+e^-$ c.m. frame, and the known mass of $D^+_s$~\cite{PDG}, respectively. $M(D^+_s\gamma)$ is the invariant mass of the $D^+_s$ candidate and a photon that minimizes the difference with $m(D^{*+}_s)$. $M(D^+_s)$ is the invariant mass of the $D^+_s$ candidate. Signals for the $D^+_s$ from $e^+e^-\to D^{*-}_s D^+_s$ form a peak structure on $M_{\rm rec}$ spectrum near $m(D_s^{*+})$, and signals for the $D^+_s$ from $e^+e^-\to D^{*+}_s D^-_s (D^{*+}_s\to \gamma D^+_s)$ process form a peak structure on $\Delta M$ spectrum around the mass difference $m_{D_s^{*+}}-m_{D^+_s}$. Here $m_{D_s^{*+}}$ is the known $D_s^{*+}$ mass~\cite{PDG}. Signal candidates must be located within the defined signal regions on the two-dimensional plane of $M_{\rm rec}$ versus $\Delta M$, which are determined independently for each energy point and decay mode based on the Punzi figure-of-merit $\epsilon/(\frac{3}{2}+\sqrt{B})$~\cite{Punzi}. The optimized requirements on $M_{\rm rec}$ and $\Delta M$ are listed in Table~\ref{Mcut}. This requirement is not applied to $D_s^+\to\phi K^-e^+e^+$ decay since its background level is already low.

\begin{table}[htbp]
\centering
\caption{Optimized requirements on $M_{\rm rec}$ and $\Delta M$ at various c.m. energies. }
\tabcolsep=0.20cm
\begin{tabular}{lccc}
\hline
    $E_{\rm cm}$~(GeV) & $M_{\rm rec}$ range~(GeV/$c^2$) & $\Delta M$ range~(GeV/$c^2$) \\
    \hline
    \multicolumn{3}{c}{$D_s^+\to K_S^0\pi^- e^+e^+$}\\
    \hline
    4.128\&4.157   & [2.1051, 2.1191]             & [0.124, 0.161] \\
    4.178          & [2.1051, 2.1201]             & [0.129, 0.150] \\
    $4.189-4.219$            & [2.1051, 2.1271]             & [0.132, 0.153] \\
    4.226          & [2.1001, 2.1251]             & [0.131, 0.149] \\
    \hline
    \multicolumn{3}{c}{$D_s^+\to K_S^0K^- e^+e^+$}\\
    \hline
    4.128\&4.157   & [2.1071, 2.1251]             & [0.124, 0.168] \\
    4.178          & [2.0871, 2.1391]             & [0.138, 0.148] \\
    $4.189-4.219$            & [2.1041, 2.1231]             & [0.124, 0.151] \\
    4.226          & [2.0821, 2.1401]             & [0.128, 0.150] \\
    \hline
    \multicolumn{3}{c}{$D_s^+\to \pi^-\pi^0 e^+e^+$}\\
    \hline
    4.128\&4.157   & [2.1031, 2.1261]             & [0.134, 0.150] \\
    4.178          & [2.1071, 2.1201]             & [0.136, 0.148] \\
    $4.189-4.219$            & [2.1031, 2.1261]             & [0.134, 0.149] \\
    4.226          & [2.1021, 2.1321]             & [0.136, 0.150] \\
    \hline
    \multicolumn{3}{c}{$D_s^+\to K^-\pi^0 e^+e^+$}\\
    \hline
    4.128\&4.157   & [2.1061, 2.1201]             & [0.136, 0.149] \\
    4.178          & [2.1051, 2.1211]             & [0.134, 0.149] \\
    $4.189-4.219$            & [2.1031, 2.1291]             & [0.132, 0.148] \\
    4.226          & [2.1041, 2.1321]             & [0.133, 0.149] \\
    \hline
    \multicolumn{3}{c}{$D_s^+\to \phi\pi^- e^+e^+$}\\\hline
    4.128\&4.157   & [2.1031, 2.1271]             & [0.137, 0.152] \\
    4.178          & [2.1051, 2.1201]             & [0.127, 0.153] \\
    $4.189-4.219$            & [2.0971, 2.1301]             & [0.137, 0.151] \\
    4.226          & [2.0881, 2.1401]             & [0.116, 0.158] \\
    \hline
    \label{Mcut}
\end{tabular}
\end{table}

\section{Signal determination}

\begin{figure*}[!]
    \centering
    \includegraphics[width=4.5cm]{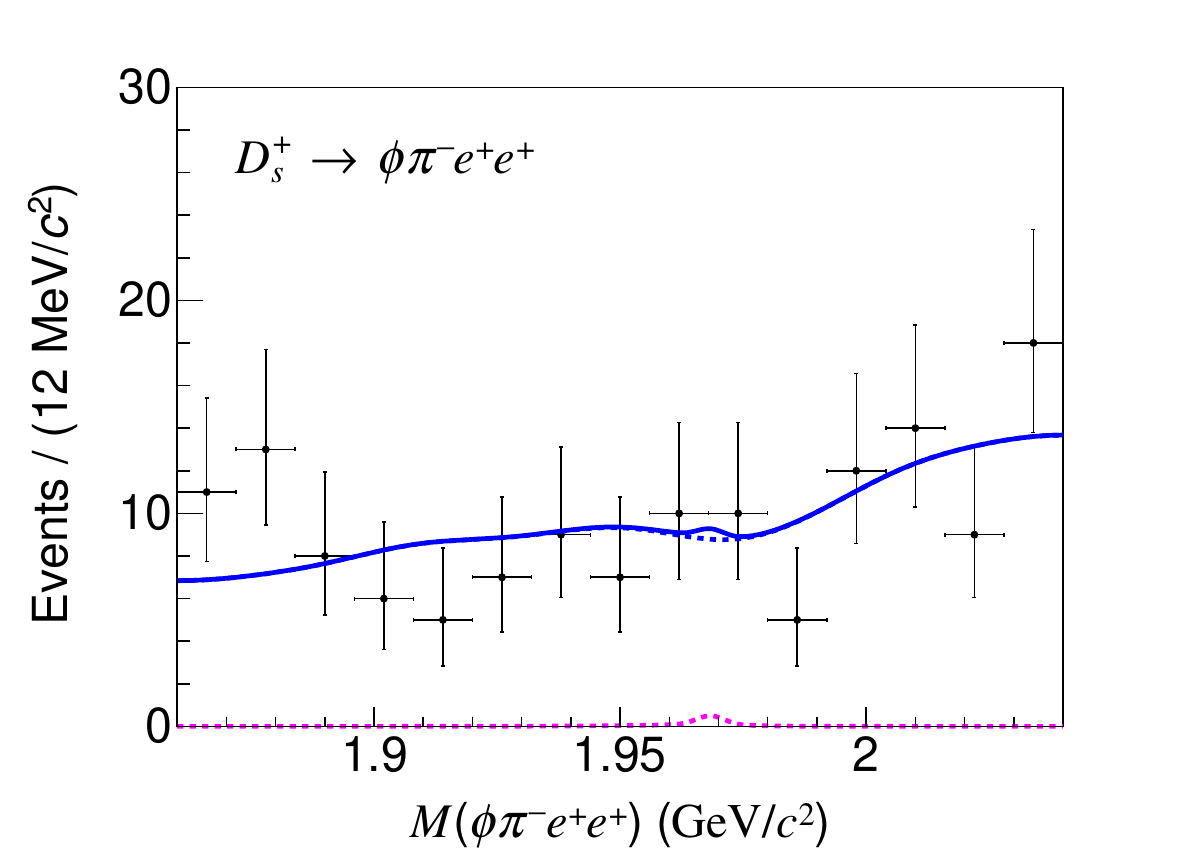}
    \includegraphics[width=4.5cm]{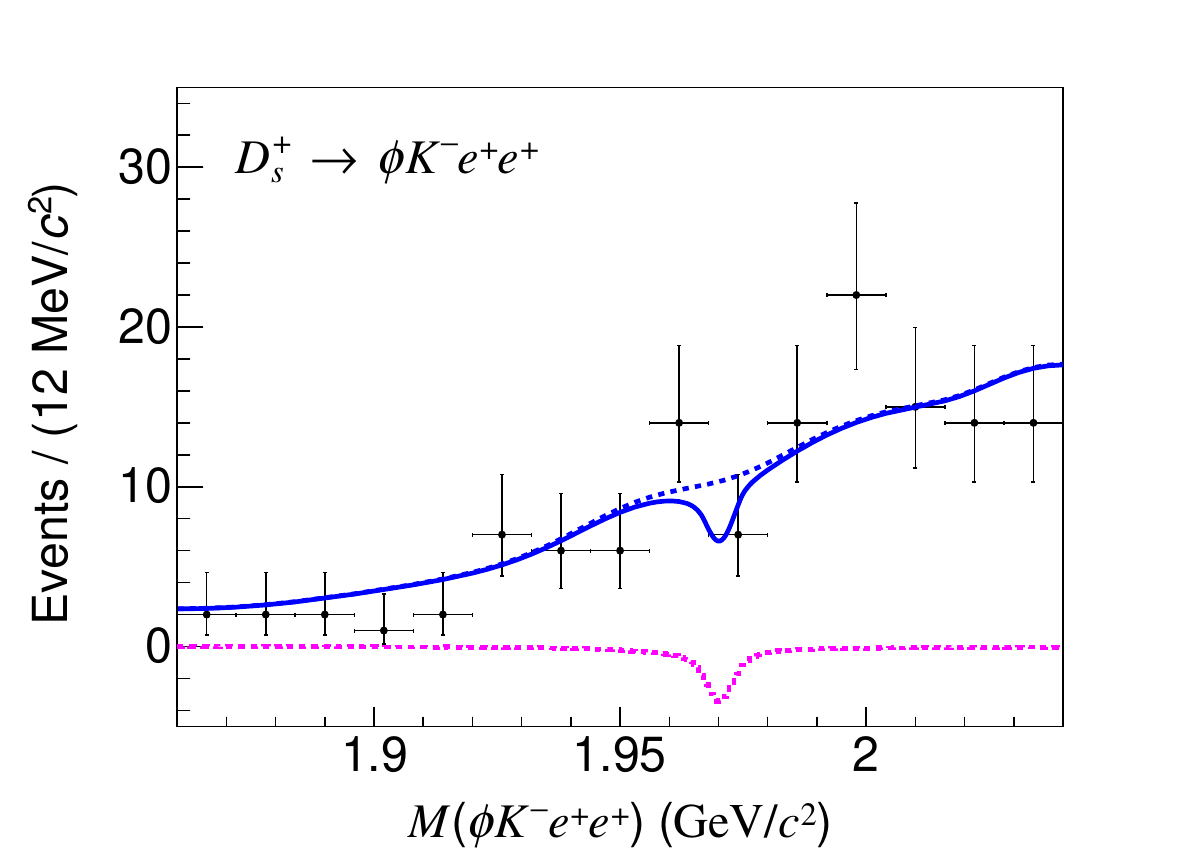}
    \includegraphics[width=4.5cm]{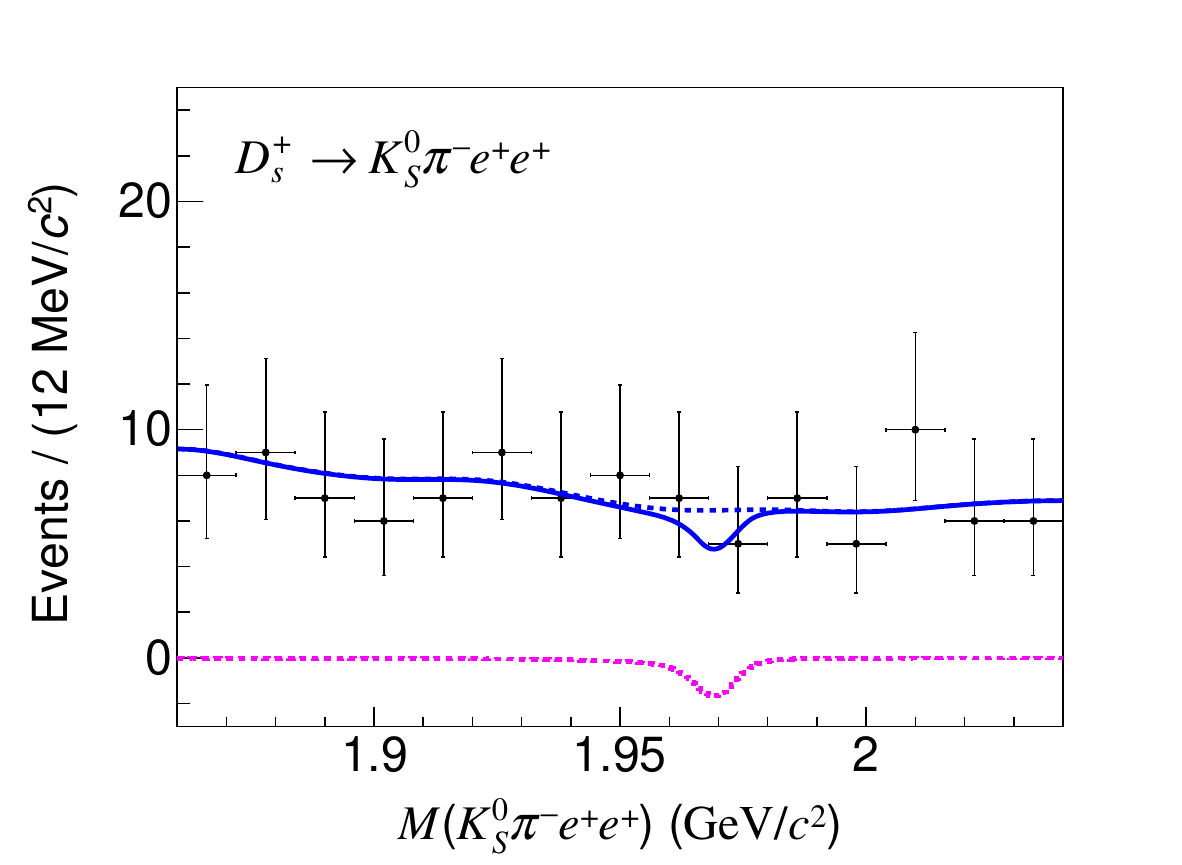}
    \includegraphics[width=4.5cm]{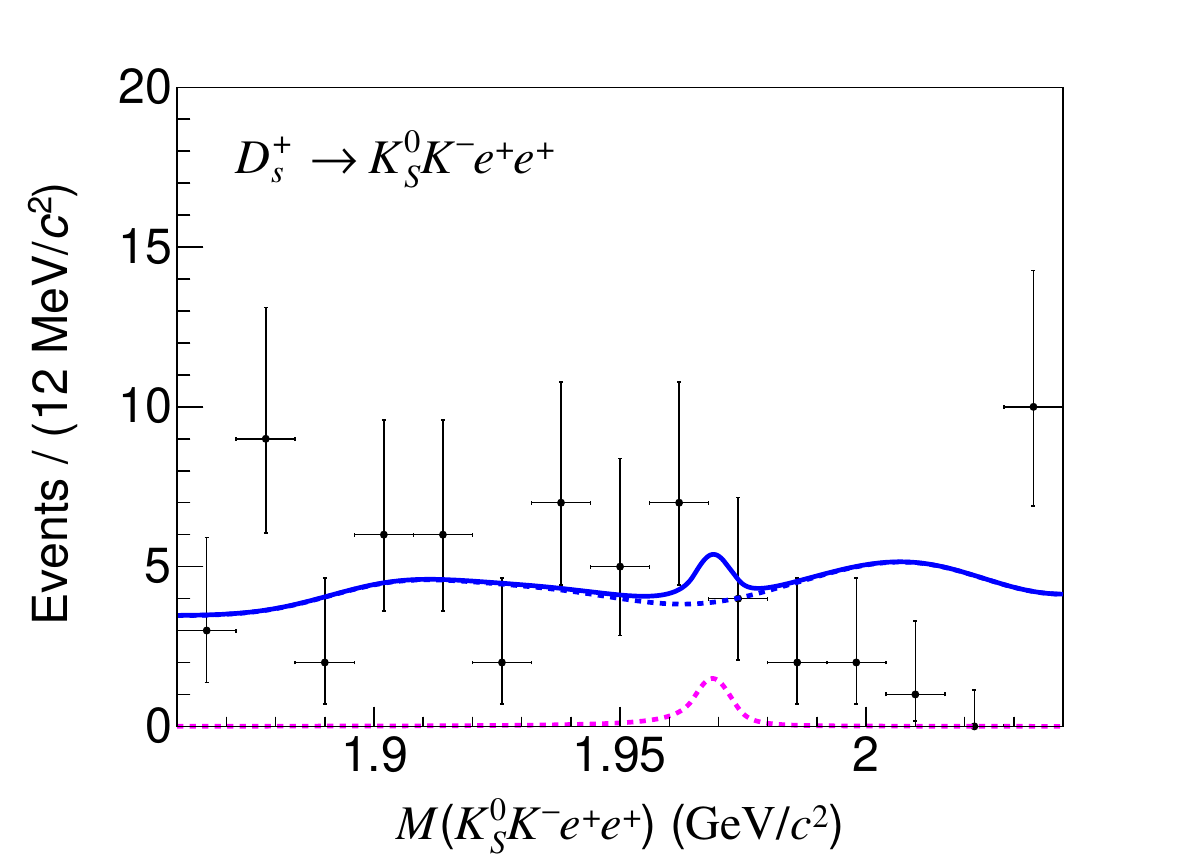}
    \includegraphics[width=4.5cm]{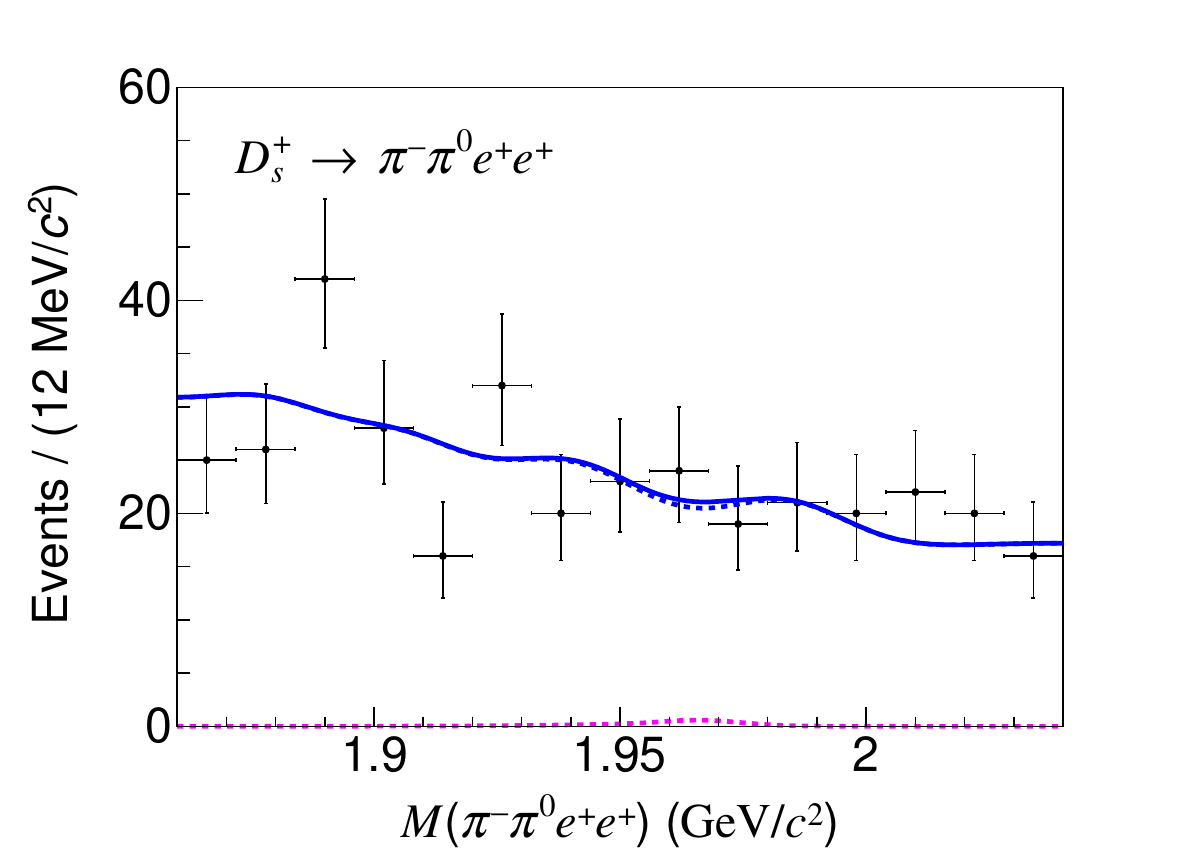}
    \includegraphics[width=4.5cm]{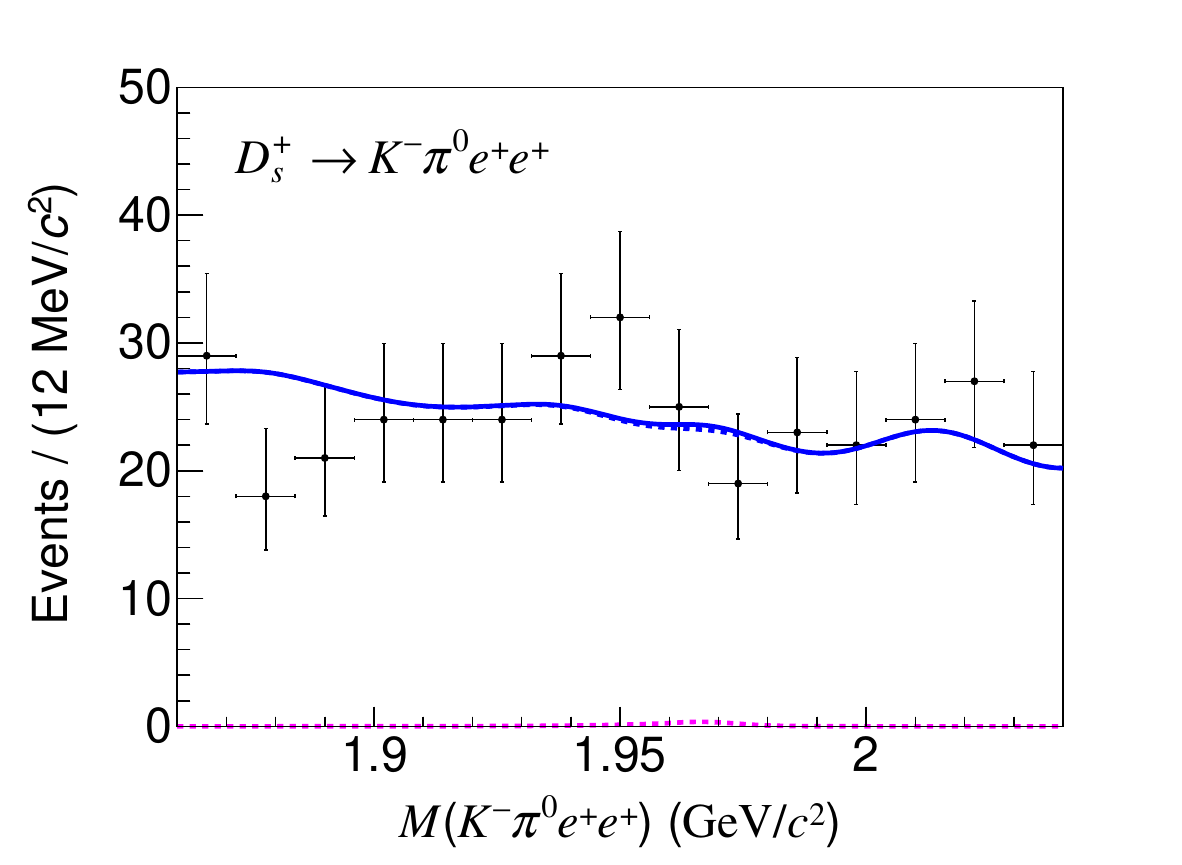}
    \caption{The fits to the invariant mass of corresponding final state of all data samples for each signal channel. The blue solid, magenta dashed and blue dashed lines represent the total fit, signal and background shapes, respectively. The downward magenta peaks denote negative signal yields.}
    \label{fitdata}
\end{figure*}

The signal yields are determined by performing an unbinned maximum likelihood fit to the invariant mass of corresponding final state ($M(h^-h^0e^+e^+)$) within the range of ${[1.86, 2.04]~\rm{GeV}/c^2}$. In the fit, the signal shape is modeled by the sum of a double-sided Crystal Ball function~\cite{crysball} and a bifurcated Gaussian function with asymmetric tails. The parameters for these functions are derived from MC samples. The background shape is modeled by inclusive MC samples using the RooKeysPdf~\cite{RooKeysPdf}. The fit results are shown in Fig.~\ref{fitdata}.

Since no obvious signal is observed, the ULs of the BFs at the 90\% C.L. for $D_s^+\to h^-h^0e^+e^+$ decays are set after accounting for systematic uncertainties.

\section{Systematic uncertainty}
    
The systematic uncertainties arise from several sources, including the tracking and PID efficiencies of charged tracks, the reconstruction of neutral particles, the total number of $D^{*\pm}_s D^\mp_s$ pairs, the BFs of intermediate state decays, the background suppression requirements, and the signal and background shapes used in the fit. All the systematic uncertainties are listed in Table~\ref{sys} and discussed below. Assuming they are independent, the total systematic uncertainty is the quadrature sum of the individual ones.

\begin{table}
  \caption{Relative systematic uncertainties (in \%) of the measurements of the ULs of the BFs of $D_s^+\to h^-h^0e^+e^+$ decays. %$K_{S}^{0}/\pi^0$ reco. represents $K_{S}^{0}/\pi^0$ reconstruction.
  }
	\centering
    \tabcolsep=0.15cm
    \begin{tabular}{lcccccc}
    \hline
    Source & $K_S^0\pi^- e^+e^+$ & $K_S^0K^- e^+e^+$ & $\pi^-\pi^0 e^+e^+$ & $K^-\pi^0 e^+e^+$ & $\phi\pi^- e^+e^+$ & $\phi K^-e^+e^+$\\
     \hline
     $\pi^{-}$ tracking                   & 0.3 & -   & 0.3 & -   & 0.3 & -\\
     $\pi^{-}$ PID                        & 0.5 & -   & 0.5 & -   & 0.5 & -\\
     $K^{\pm}$ tracking                     & -   & 0.8 & -   & 0.8 & 1.6 & 2.4\\
     $K^{\pm}$ PID                          & -   & 0.8 & -   & 0.8 & 1.6 & 2.4\\
     $e^{+}$ tracking                     & 1.3 & 1.8 & 1.1 & 1.4 & 1.7 & 1.7\\
     $e^{+}$ PID                          & 1.1 & 1.2 & 1.1 & 1.2 & 1.6 & 2.2\\
     $K_{S}^{0}/\pi^0$ reco. & 1.5 & 1.5 & 2.0 & 2.0 & -   & -\\
     $N_{D_s}^{\rm{tot}}$                  & 0.5 & 0.5 & 0.5 & 0.5 & 0.5 & 0.5\\
     $\mathcal{B}_{\rm{inter}}$            & -   & -   & -   & -   & 1.0 & 1.0\\ 
     $M_{\rm{rec}} \& \Delta M$            & 4.0 & 5.6 & 9.5 & 9.8 & 8.8 & -\\
     \hline
     Total                           & 5.4 & 7.4 & 10.2 & 10.9 & 11.5 & 8.8\\
     \hline\\
     \label{sys}
\end{tabular}
\end{table}

The uncertainties from the tracking (PID) of $K^\pm$ and $\pi^-$ are assigned to be 0.8\% (0.8\%) and 0.3\% (0.5\%) per track, studied using samples of \mbox{$e^+e^-\to K^+K^-\pi^+\pi^-$} [\mbox{$K^+K^-K^+K^-$}, \mbox{$K^+K^-\pi^+\pi^-(\pi^0)$} and \mbox{$\pi^+\pi^-\pi^+\pi^-(\pi^0)$}] events~\cite{sys_kpiks}. The uncertainties associated with the tracking and PID efficiencies of $e^+$ are studied with a control sample of radiative Bhabha process($e^+e^-\to \gamma e^+e^-$) ~\cite{sys_e}. Signal MC events are weighted event-by-event with a correction factor, $\frac{\epsilon_{Data}}{\epsilon_{MC}}$, to account for discrepancies in tracking and PID efficiencies between data and MC simulations as a function of electron momentum and track angle. The relative differences between tracking (PID) efficiencies before and after the weighting process are taken as the systematic uncertainties. The uncertainties from $\pi^0$ reconstruction is assigned as 2.0\% per $\pi^0$, based on a control sample of $e^+e^-\to K^+K^-\pi^+\pi^-\pi^0$~\cite{sys_pi0}. The uncertainties from $K^0_S$ reconstruction is assigned as 1.5\% per $K^0_S$, based on control samples of $J/\psi\to K_S^0K^{\pm}\pi^{\mp}$ and $J/\psi\to \phi K_S^0K^{\pm}\pi^{\mp}$ decays~\cite{sys_kpiks}. 

The uncertainty of the total number of $D^{*\pm}_s D^\mp_s$ pairs is 0.4\%~\cite{Ds_number}. The possible contamination from $e^+e^-\to D^+_sD^-_s$ is estimated to be 0.3\%. Therefore, the total systematic uncertainty from the number of $D^{*\pm}_s D^\mp_s$ pairs is assigned to be 0.5\%. The uncertainties from the BFs of $K_{S}^{0}\to \pi^+\pi^-$ and $\pi^0\to\gamma\gamma$ decays are negligible, while the uncertainty from the BF of $\phi\to K^{+}K^{-}$ decay is 1\% according to the PDG~\cite{PDG}. 

The systematic uncertainty from the requirements on $M_{\rm rec}$ and $\Delta M$ is studied using the ST control samples of $D_s^+\to K_S^0K^-\pi^+\pi^+$, $D_s^+\to K^+K^-\pi^+\pi^0$ and $D_s^+\to K^+K^-\pi^+\pi^-\pi^+$ decays. The relative difference in efficiencies between data and MC samples with and without the requirements is treated as the systematic uncertainty. The systematic uncertainties from the requirements on cos($\theta_{e^+e^-}$), cos($\theta_{e^+\pi^-}$), $L_\pi/\sigma_{L_\pi}$, $E(\pi^0)$, $L/\sigma_L$, $E/p$ and $\chi_{{\rm d}E/{\rm d}x}^2$, are examined by varying individual requirement by $\pm0.02$, $\pm0.02$, $\pm1$, ${\pm0.01~\rm{GeV}}$, $\pm0.5$, $\pm0.04c$ and $\pm0.3$, respectively. Accounting for correlations in the samples, the changes in signal yields are smaller than the statistical uncertainty on the difference, so the systematic uncertainties from these requirements are negligible according to Ref.~\cite{Barlow}. 

The systematic uncertainty from the signal shape used in the fit is evaluated by performing an alternative fit with the signal described by the shape from MC simulation convolved with a Gaussian function. The changes in signal yields are smaller than the statistical uncertainty on the difference, indicating that the systematic uncertainties from these requirements are negligible~\cite{Barlow}. The systematic uncertainty from the background shape is studied by varying the background model from the shape derived from the inclusive MC sample to a $2^{\rm{nd}}$ order polynomial function. The larger UL is adopted as the final result.

\section{Results}
\subsection{Upper limits of branching fractions of $D_s^+\to h^-h^0e^+e^+$ decays}
\begin{figure}
    \centering
    \includegraphics[width=4.5cm]{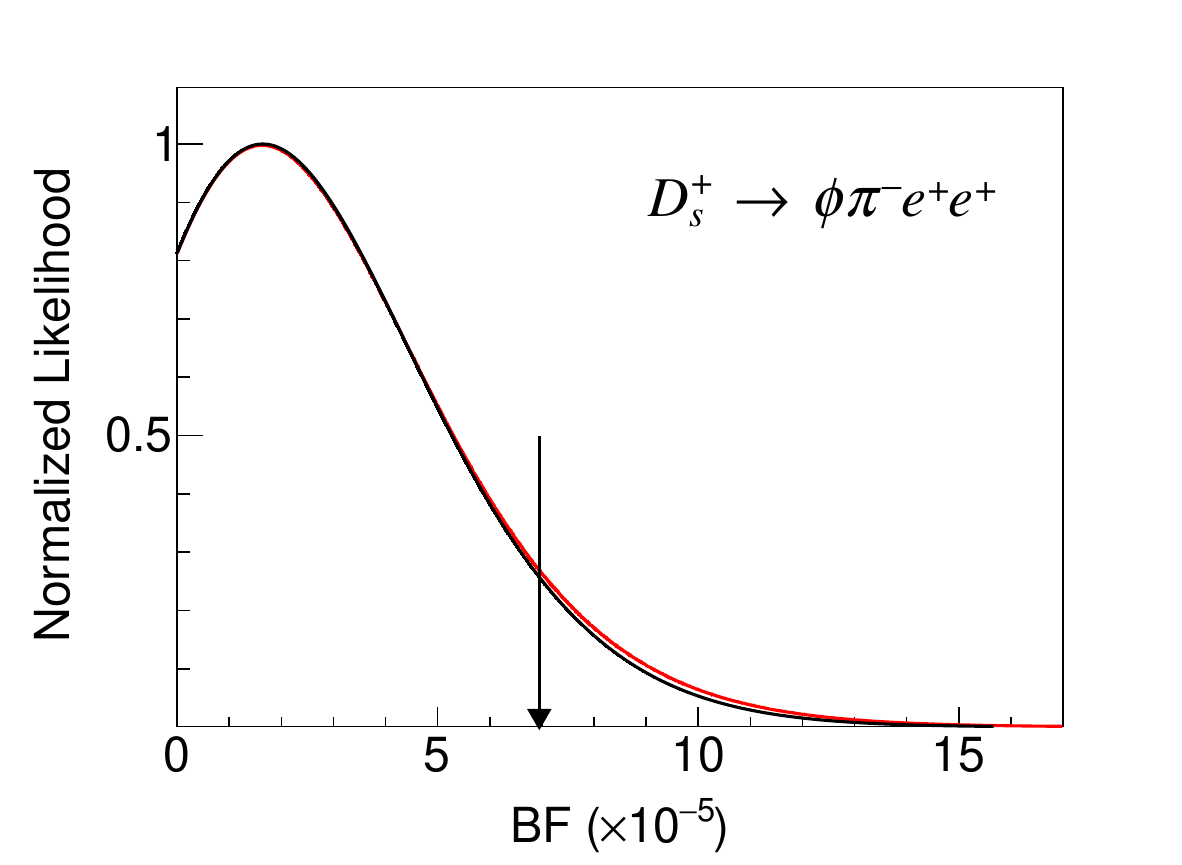}
    \includegraphics[width=4.5cm]{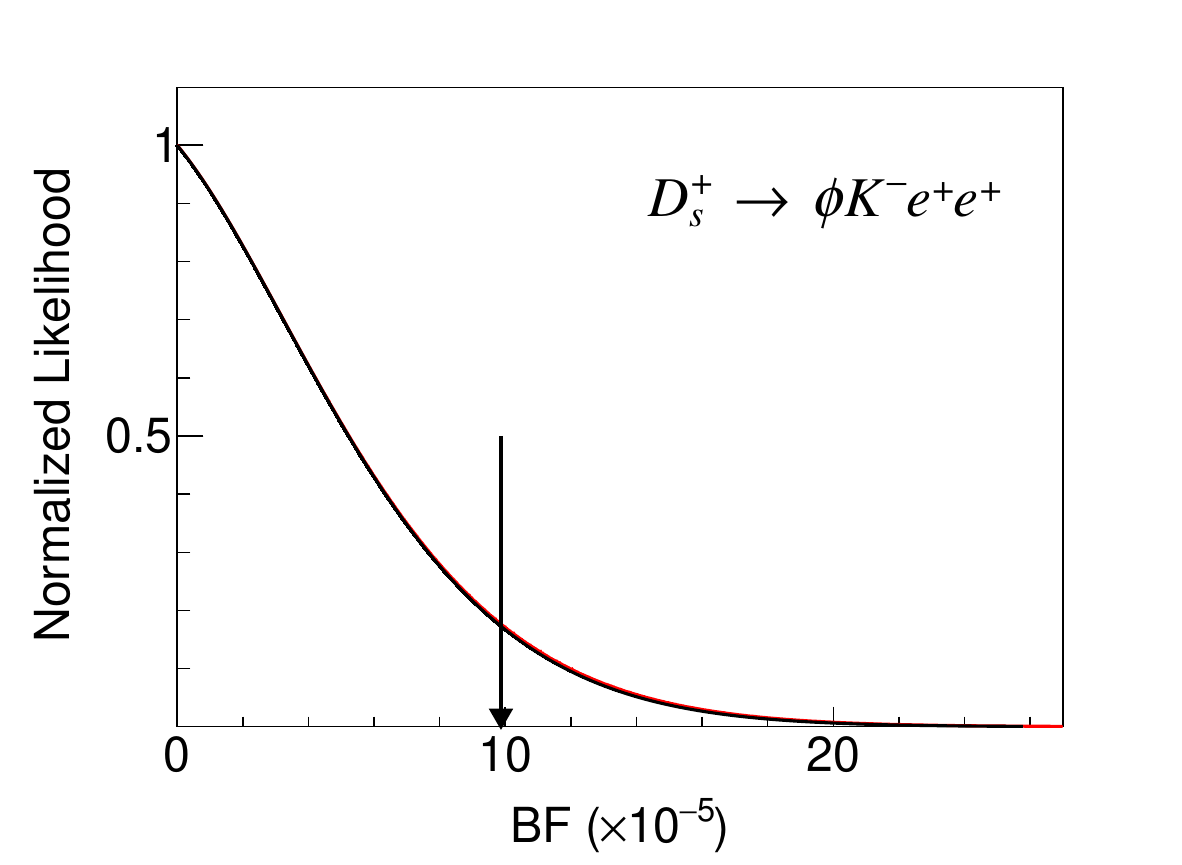}
    \includegraphics[width=4.5cm]{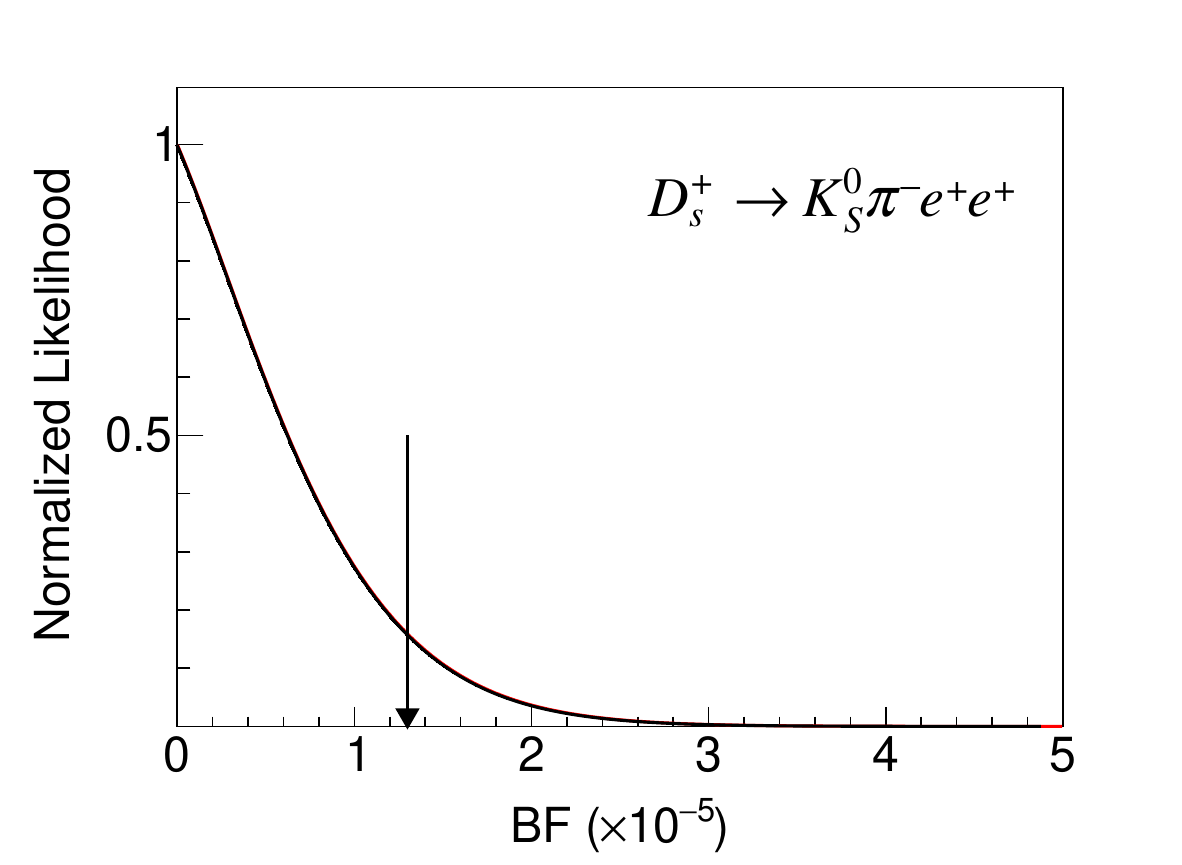}
    \includegraphics[width=4.5cm]{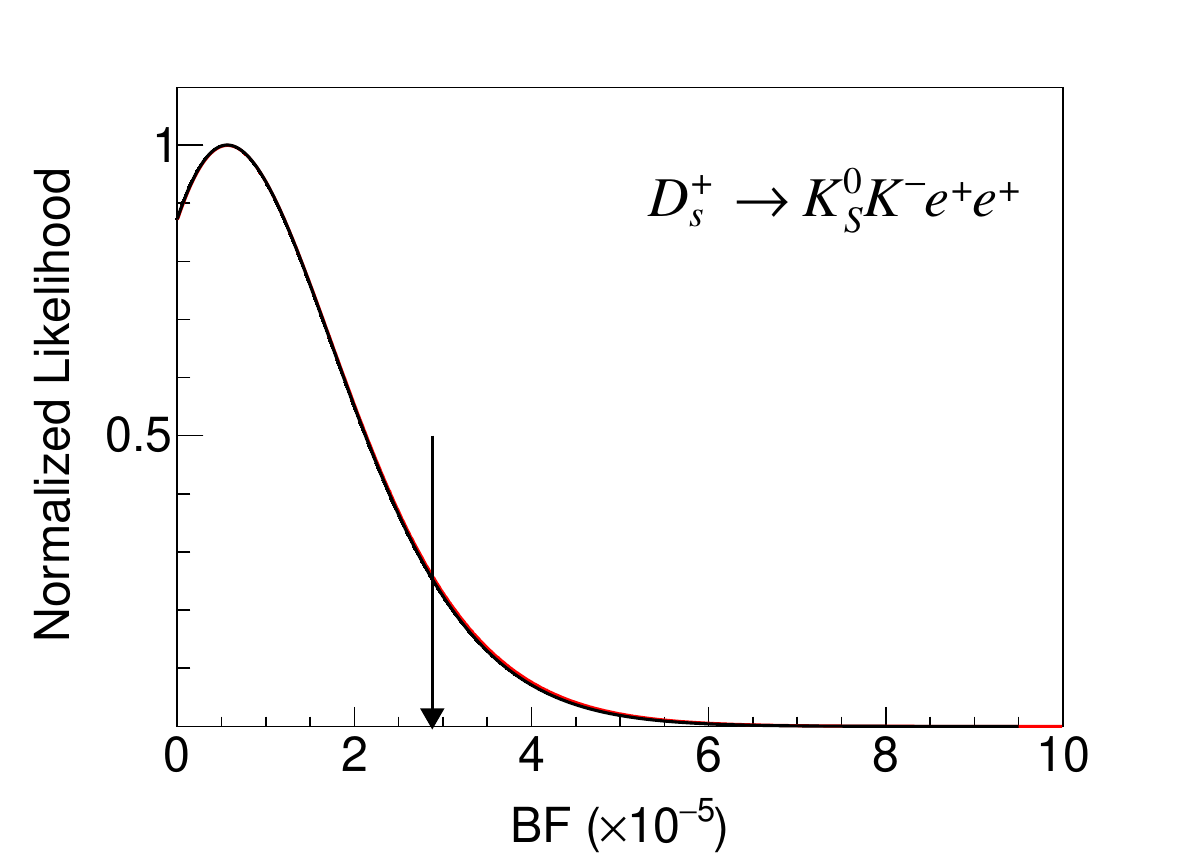}
    \includegraphics[width=4.5cm]{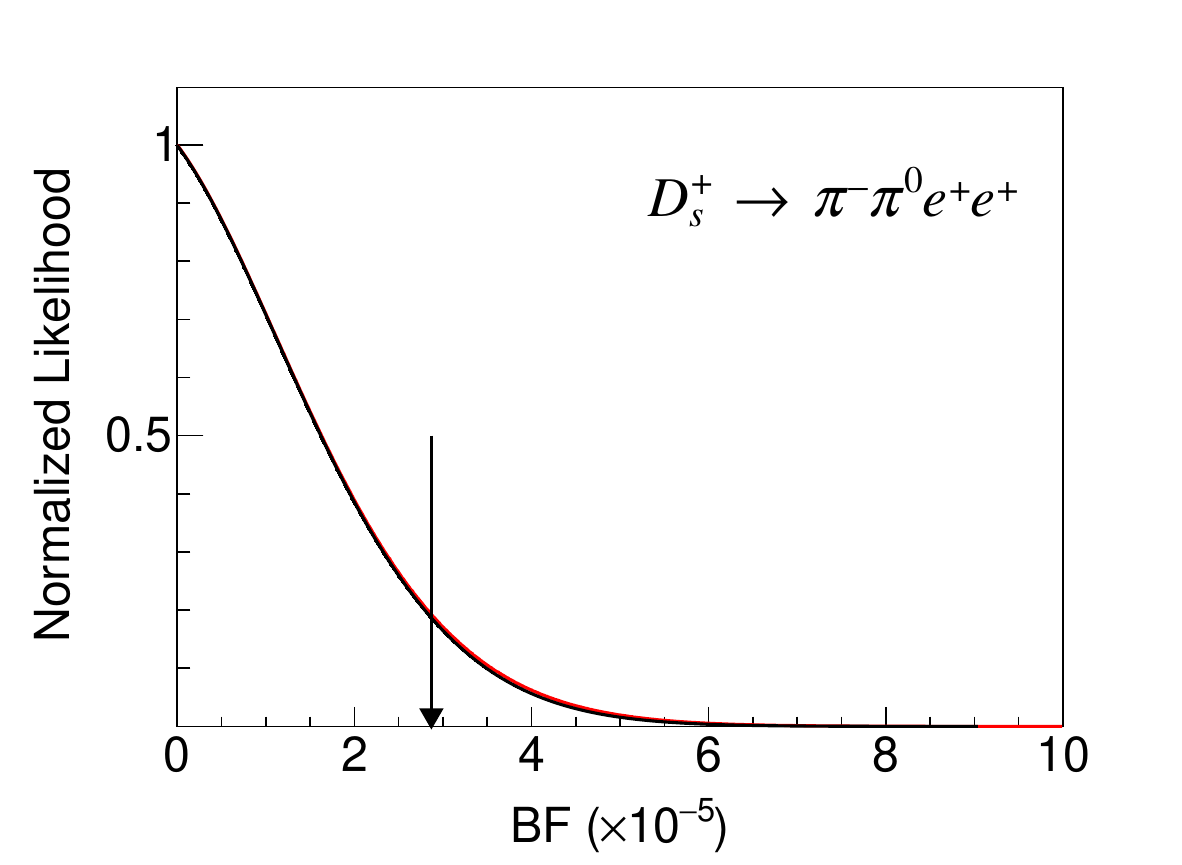}
    \includegraphics[width=4.5cm]{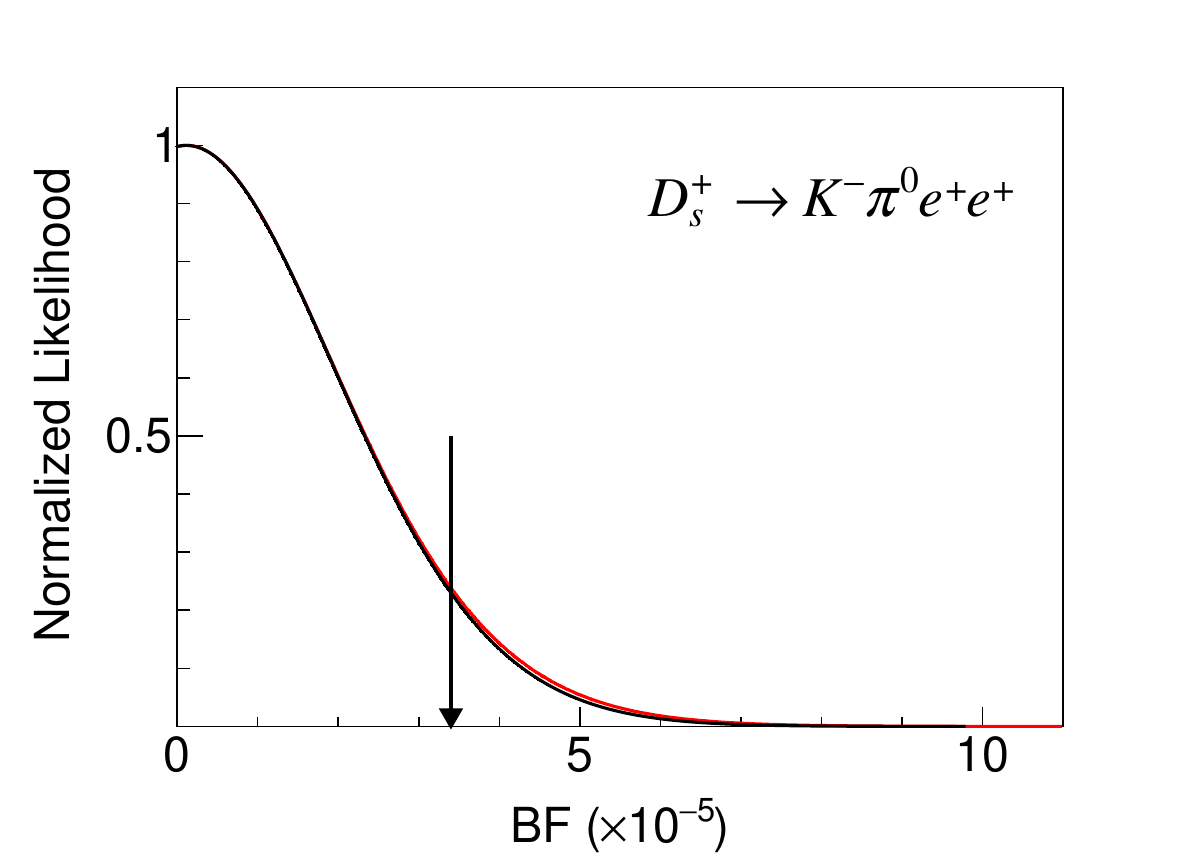}
    \caption{The likelihood distributions depending on the corresponding BF for each signal channel, with (red solid curves) and without (black solid curves) incorporating the systematic uncertainties. The black arrows denote the ULs of the BFs at the 90\% C.L.}
    \label{llcurve}
\end{figure}

Using a Bayesian method~\cite{ULscan}, a likelihood scan is performed by fixing the signal yield at various values. Systematic uncertainties, excluding those associated with the background shape (as discussed in the above section), are incorporated by convolving the likelihood curve with a Gaussian function, where the standard deviation of the Gaussian function represents the total systematic uncertainty. The normalized likelihood distributions as a function of BF for each signal channel are shown in Fig.~\ref{llcurve}. Based on a background-only hypothesis, expected ULs of the BFs are determined. Both expected and observed ULs of the BFs at the 90\% C.L. are summarized in Table~\ref{UL}. A frequentist technique is applied to measure the ULs of the BFs as a cross-check~\cite{belleBNV}, and the results are consistent with those derived from the Bayesian approach.

\begin{table}[htb]
\centering
\caption{The detection efficiencies~($\epsilon$), observed and expected ULs of the BFs at the 90\% C.L.~($\mathcal{B}_{\rm{UL}}$ and $\mathcal{B}_{\rm{UL}}^{\rm expected}$) of $D_s^+\to h^-h^0e^+e^+$ decays. No signal is assumed for the expected upper limits. The uncertainties of the detection efficiencies are statistical uncertainties. }
\vspace{0.1cm}
\label{UL}
\tabcolsep=0.3cm
\begin{tabular}{lcc}
\hline
Decay channel                & $\epsilon~(\%)$ & $\mathcal{B}_{\rm{UL}}~(\mathcal{B}_{\rm{UL}}^{\rm expected})$\\\hline
$D_s^+\to \phi\pi^- e^+e^+$  & $3.0\pm0.1$            & $6.9~(3.5)\times 10^{-5}$\\
$D_s^+\to \phi K^-e^+e^+$    & $1.8\pm0.1$            & $9.9~(10.8)\times 10^{-5}$\\
$D_s^+\to K_S^0\pi^- e^+e^+$ & $6.4\pm0.1$            & $1.3~(2.4)\times 10^{-5}$\\
$D_s^+\to K_S^0K^- e^+e^+$   & $4.0\pm0.1$            & $2.9~(2.3)\times 10^{-5}$\\
$D_s^+\to \pi^-\pi^0 e^+e^+$ & $6.4\pm0.1$            & $2.9~(2.7)\times 10^{-5}$\\
$D_s^+\to K^-\pi^0 e^+e^+$   & $5.1\pm0.1$            & $3.4~(3.9)\times 10^{-5}$\\
\hline
\end{tabular}
\end{table}

\subsection{Search for Majorana neutrino in $D^+_s\to \phi\pi^-e^+e^+$ decay}
\begin{figure}[htb]
    \centering
    \includegraphics[width=7.0cm]{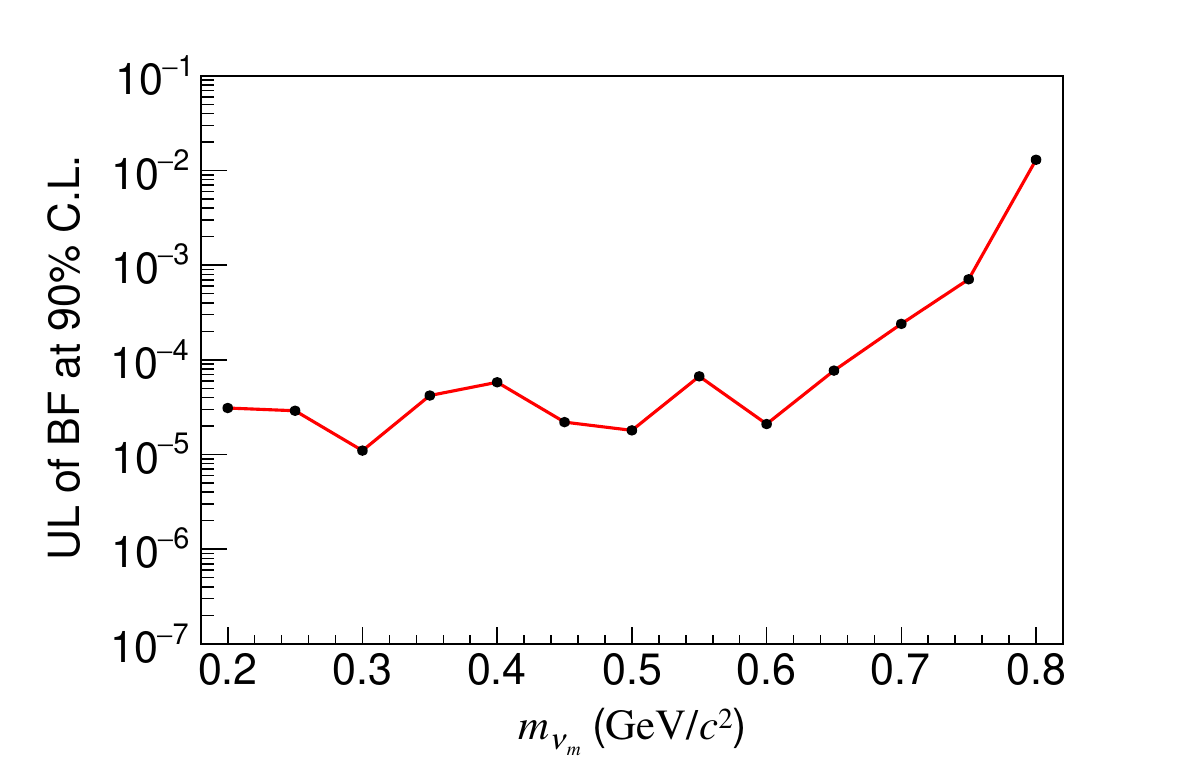}
    \caption{The ULs of the BFs at the 90\% C.L. as a function of $m_{\nu_m}$ for the $D_s^+\to\phi e^+\nu_m(\to\pi^-e^+)$ decay.}
    \label{nv_UL}
\end{figure}

Furthermore, the Majorana neutrino is searched for in the decay of $D_s^+\to\phi e^+\nu_m(\to\pi^-e^+)$ with various $m_{\nu_m}$ assumptions, ranging from 0.20 to ${0.80~\rm{GeV}/c^2}$ in intervals of 0.05 GeV/$c^2$. To search for the Majorana neutrino with a given mass, $m_{\nu_m}$, the candidate events are selected by further requiring the invariant mass of any $\pi^-e^+$ combination (two $e^+$s per event), $M_{\pi^-e^+}$, to be within the range of $[m_{\nu_m} - 5\sigma, m_{\nu_m} + 4\sigma]$, where $\sigma$ is the resolution of the $M_{\pi^-e^+}$ distribution obtained by MC simulation. Given the small number of events that pass all selection criteria, the number of signal and background events are obtained by counting. The numbers of signal candidates within the $M(\phi\pi^-e^+e^+)$ signal region, defined as $[1.940, 1.984]$ GeV$/c^2$, and the number of background events in the $M(\phi\pi^-e^+e^+)$ sideband regions, defined as $[1.860, 1.940]$ GeV$c^2$ and $[1.984,2.040]$ GeV$/c^2$, are obtained by counting. The ULs of the BFs are then calculated using the profile likelihood method taking into consideration the systematic uncertainty with the {\sc trolke}~\cite{frequentist, TROLKE} package in the {\sc root} framework. In the calculation of the ULs, the numbers of signal and background events are assumed to follow a Poisson distribution, while the detection efficiency is assumed to follow a Gaussian distribution. The ULs of the BFs at the 90\% C.L., as a function of $m_{\nu_m}$, range from around $10^{-5}-10^{-2}$, as shown in Fig.~\ref{nv_UL}.

\section{Summary}
Using 7.33 fb$^{-1}$ of $e^+e^-$ collision data taken at c.m. energies from 4.128 to 4.226 GeV, a search for LNV (${\Delta L=2}$) decays of $D_s^+\to h^-h^0e^+e^+$ is performed. Additionally, the Majorana neutrino is searched for in $D^+_s\to \phi\pi^-e^+e^+$ with various mass assumptions. No significant signal is observed. With the Bayesian approach, the ULs of the BFs at the 90\% C.L. are set to be $\mathcal{B}(D_s^+\to \phi\pi^-e^+e^+) < 6.9 \times 10^{-5}$, $\mathcal{B}(D_s^+\to \phi K^-e^+e^+) < 9.9 \times 10^{-5}$, $\mathcal{B}(D_s^+\to K_S^0\pi^-e^+e^+) < 1.3 \times 10^{-5}$, $\mathcal{B}(D_s^+\to K_S^0K^-e^+e^+) < 2.9 \times 10^{-5}$, $\mathcal{B}(D_s^+\to \pi^-\pi^0e^+e^+) < 2.9 \times 10^{-5}$ and $\mathcal{B}(D_s^+\to K^-\pi^0e^+e^+) < 3.4 \times 10^{-5}$. Furthermore, for the decay $D_s^+\to\phi e^+\nu_m(\to\pi^-e^+)$, the ULs of the BFs at the 90\% C.L. with different $m_{\nu_m}$ assumptions in the range of ${[0.20, 0.80]~\rm{GeV}/c^2}$ are also determined, which are at the level of $10^{-5}-10^{-2}$.

% \end{linenumbers}
\acknowledgments
The BESIII Collaboration thanks the staff of BEPCII and the IHEP computing center for their strong support. This work is supported in part by National Key R\&D Program of China under Contracts Nos. 2023YFA1606000, 2020YFA0406300, 2020YFA0406400; National Natural Science Foundation of China (NSFC) under Contracts Nos. 11635010, 11735014, 11935015, 11935016, 11935018, 12025502, 12035009, 12035013, 12061131003, 12192260, 12192261, 12192262, 12192263, 12192264, 12192265, 12221005, 12225509, 12235017, 12361141819; the Chinese Academy of Sciences (CAS) Large-Scale Scientific Facility Program; the CAS Center for Excellence in Particle Physics (CCEPP); Joint Large-Scale Scientific Facility Funds of the NSFC and CAS under Contract No. U1832207; 100 Talents Program of CAS; The Institute of Nuclear and Particle Physics (INPAC) and Shanghai Key Laboratory for Particle Physics and Cosmology; German Research Foundation DFG under Contracts Nos. FOR5327, GRK 2149; Istituto Nazionale di Fisica Nucleare, Italy; Knut and Alice Wallenberg Foundation under Contracts Nos. 2021.0174, 2021.0299; Ministry of Development of Turkey under Contract No. DPT2006K-120470; National Research Foundation of Korea under Contract No. NRF-2022R1A2C1092335; National Science and Technology fund of Mongolia; National Science Research and Innovation Fund (NSRF) via the Program Management Unit for Human Resources \& Institutional Development, Research and Innovation of Thailand under Contracts Nos. B16F640076, B50G670107; Polish National Science Centre under Contract No. 2019/35/O/ST2/02907; Swedish Research Council under Contract No. 2019.04595; The Swedish Foundation for International Cooperation in Research and Higher Education under Contract No. CH2018-7756; U. S. Department of Energy under Contract No. DE-FG02-05ER41374.

\clearpage
%\author{Author list}
%\begin{small}
%\begin{center}
\large
The BESIII Collaboration\\
\normalsize
%% Saved at => 2023-03-31
\\ M.~Ablikim$^{1}$, M.~N.~Achasov$^{4,c}$, P.~Adlarson$^{76}$, O.~Afedulidis$^{3}$, X.~C.~Ai$^{81}$, R.~Aliberti$^{35}$, A.~Amoroso$^{75A,75C}$, Q.~An$^{72,58,a}$, Y.~Bai$^{57}$, O.~Bakina$^{36}$
% ...
\\
\vspace{0.2cm}
(BESIII Collaboration)\\
\vspace{0.2cm} {\it
$^{1}$ Institute of High Energy Physics, Beijing 100049, People's Republic of China\\
% ...
$^{3}$ Bochum Ruhr-University, D-44780 Bochum, Germany\\
$^{4}$ Budker Institute of Nuclear Physics SB RAS (BINP), Novosibirsk 630090, Russia\\
% ...
$^{35}$ Johannes Gutenberg University of Mainz, Johann-Joachim-Becher-Weg 45, D-55099 Mainz, Germany\\
$^{36}$ Joint Institute for Nuclear Research, 141980 Dubna, Moscow region, Russia\\
% ...
$^{57}$ Southeast University, Nanjing 211100, People's Republic of China\\
$^{58}$ State Key Laboratory of Particle Detection and Electronics, Beijing 100049, Hefei 230026, People's Republic of China\\
% ...
$^{72}$ University of Science and Technology of China, Hefei 230026, People's Republic of China\\
% ...
$^{75}$ University of Turin and INFN, (A)University of Turin, I-10125, Turin, Italy; (B)University of Eastern Piedmont, I-15121, Alessandria, Italy; (C)INFN, I-10125, Turin, Italy\\
$^{76}$ Uppsala University, Box 516, SE-75120 Uppsala, Sweden\\
% ...
$^{81}$ Zhengzhou University, Zhengzhou 450001, People's Republic of China\\
\vspace{0.2cm}
$^{a}$ Deceased\\
% ...
$^{c}$ Also at the Novosibirsk State University, Novosibirsk, 630090, Russia\\
% ...
}
%% ends here %%
%\end{center}
%\vspace{0.4cm}
%\end{small}

\end{document}